# Concept framework for the safety assessment of platooning trucks enabled by V2V communication*

Olaf Op den Camp, Jacco van de Sluis

*Abstract* — Large steps are being taken by the industry and R&D organizations in automated driving technology development, as well as in setting up appropriate scenario-based safety assessment methods. In this paper a concept framework is proposed for a very specific group of vehicles: trucks in a platoon enabled by vehicle-to-vehicle (V2V) communication. The basis is formed by previously proposed scenario-based assessment methods for individual automated driving vehicles. These methods however do not consider the inter-vehicle communication component. It is shown how V2V communication interconnects vehicles in a system-of-systems, and how to include V2V communication in the scenario description. The paper also provides a vision to answer the basic question whether safety assessment should consider the platoon as a whole, or the individual vehicle in the platoon.

## I. INTRODUCTION

The effort that is put into the development of automated vehicles by the industry is unmistakably very high. In the automation challenge, we distinguish between performance of perception, how well does the sensor system detect, identify, track and predict behavior of objects possibly interfering with the ego vehicle. And between the performance of control & decision logic, how well is the ego vehicle capable of anticipating on this behavior as early as possible in the best possible way. Of course, if proper perception of the sensor system fails, then anticipation becomes very difficult or even impossible. The development of a fair and reliable safety assessment framework is important for the safe deployment of automated vehicles on the public road, i.e. to test performance of perception and performance of control & decision logic. Results are important for authorities to monitor the safety of vehicles that they allow on the road and to steer policy with regard to implementation of automated vehicles, and for the industry, to get an understanding how their automated vehicle performs in terms of safety on the road, as early as in the development phase.

Safety assessment frameworks that are based on real-world scenarios are considered to be a structured way of dealing with the infinite different situations that an automated vehicle needs to be able to deal with in a safe way when deployed on the public road [1, 2, 3, 4]. So far, safety assessment frameworks consider single automated vehicles that base their responses on their view at the surrounding traffic and its environment. Sensor systems based on radar, lidar and/or camera techniques collect that view, where sensor fusion on board of each individual automated vehicle is used to build a single world model. The world model is the most important input to the automated vehicle's decision and control logic, in order for the vehicle to provide an appropriate response.

An important technology to enable higher levels of automation is V2X communication, which considers both the information exchange between vehicles and the infrastructure (V2I) and the information exchange between vehicles through V2V communication. The latter, V2V communication, is an indispensable technology to enable safe platooning of trucks. Platooning of long-haul trucks is expected to show large benefits regarding reductions in fuel consumption [5], improvements of traffic throughput as a result of more efficient utilization of road capacity and improved road safety by decreasing the role of the human in controlling the vehicle [6]. Regarding fuel consumption and traffic throughput, results improve with decreasing time-headway (THW[†]) between the trucks in the platoon. Conventional Advanced Cruise Control (ACC) systems, measuring relative velocity and distance with respect to the preceding vehicle in the platoon using sensors such as camera, radar and/or lidar, are not able to assure string stability in the platoon unless relatively large intervehicle distances are chosen [7]. With V2V communication, string stability in the platoon can be achieved even for short following distances.

In the European Horizon 2020 project ENSEMBLE [8], technology is developed to demonstrate heterogeneous multi-brand truck platooning enabled by V2V communication. As current frameworks for safety assessment of (highly) automated vehicles do not take inter-vehicle communication (V2V) or more generic V2X into account, it is not possible to simply apply such a safety assessment framework to the use case of truck platooning. The European Horizon 2020 project HEADSTART [9] aims to define testing and validation procedures of Connected and Automated Driving (CAD) functions and in this way supports the safety assessment of the truck platooning solutions developed in ENSEMBLE. This paper shows how the definition of the concept 'scenario' [10] is extended for applications making use of V2V communication and how safety assessment of a platoon of trucks may look like based on the extended scenario definition.

* This work is part of the HEADSTART project. This project has received funding from the European Union's Horizon 2020 research and innovation programme under grant agreement No 824309. Content reflects only the authors' view and European Commission is not responsible for any use that may be made of the information it contains.

Olaf Op den Camp (corresponding author) and Jacco van de Sluis are with TNO Integrated Vehicle Safety, the Netherlands (phone: +31 88 866 5408; e-mail: olaf.opdencamp@tno.nl).

[†] THW is defined as the distance between the rear of the preceding truck and the front of the following truck, divided by the vehicle speed.

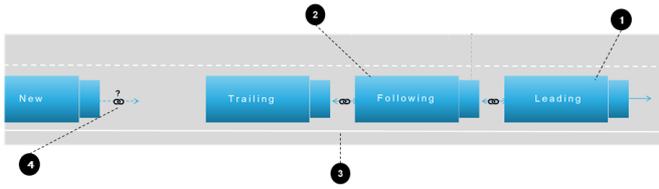

Fig. 1: Level A platooning according to the definitions from the ENSEMBLE project. 1) the leading truck is manually driven by a human driver; 2) the following truck has longitudinal automation, no lateral automation, hence the following truck are steered by human drivers; 3) the THW ≥ 0.8 s during platooning; 4) new candidate truck can only engage the platoon from the rear.

## II. PROBLEM FORMULATION

For performing a vehicle safety assessment, in the HEADSTART project [9] test cases will be determined for different use cases that specifically address Key Enabling Technologies (KETs) important for introducing Connected Cooperative Automated Driving at a large scale. Vehicle-to-Everything (V2X) communication is one of these KETs. ENSEMBLE truck platooning is selected as use case for HEADSTART, as V2V communication is key to enable platooning between vehicles (see Fig. 1).

To arrive at appropriate and relevant test cases, a scenario-based approach [11] is followed. A scenario gives a description of a situation that a vehicle might encounter during its lifetime on the road and to which the vehicle needs to respond appropriately:

**Definition** — *A scenario is a quantitative description of the ego vehicle, its activities and/or goals, its dynamic environment (consisting of the traffic environment and the light & weather conditions) and its static environment. From the perspective of the ego vehicle, a scenario contains all relevant events.*

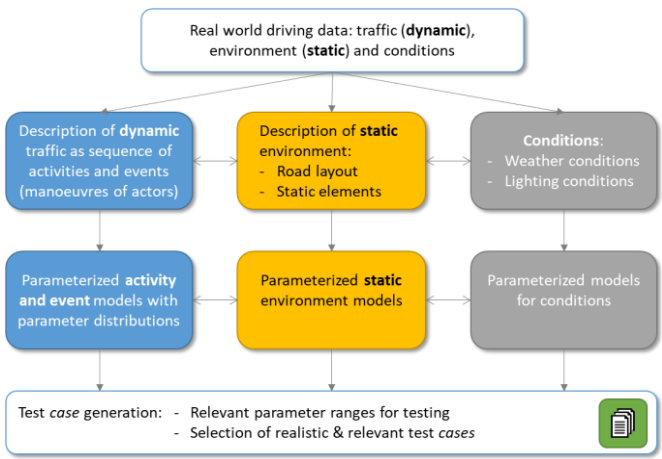

Fig. 2: Schematic view on the definition of *scenario* and the relation between scenarios and test case generation according to *[11]*.

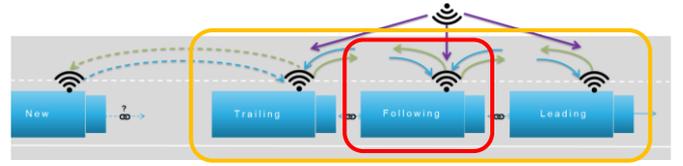

Fig. 3: Two essentially different ways of considering system boundaries for the platooning use case. Where the orange box considers the platoon as the system (of systems), the red box considers the individual vehicle.

In this scenario definition, the role of communication between vehicles, important in a platoon for sharing intentions and to "extend" the sensor range, has not yet been considered. In addition, platoon actions are coordinated amongst the vehicles of the platoon, so received inputs can influence the outputs of V2V. In the next section, first the concept of platooning will be further elaborated before studying appropriate scenario extensions for safety assessment.

## III. PLATOONING

In the ENSEMBLE project, platoon levels [8] are proposed showing the level of automation of a platoon. In this paper, we focus on platooning Level A, as this is the first level of automation that will come to the road. If the extension of the scenario description works for Level A Platoons, then it is expected to work for the other levels as well. In a Level A Platoon, the first truck in the platoon is driven manually by a human driver, possibly supported by advanced driver assistance systems (ADAS). The following trucks have fully automated longitudinal control, whereas a (safety) driver in each of these trucks is responsible to keep the following truck in its lane, though also such lateral control of the truck might be supported with an ADAS system. The time headway between trucks is minimal 0.8 seconds, and can be (automatically) adapted for local situations, e.g. to 1,5 seconds for downhill conditions, or manually adjusted to driver preferences. New candidate trucks can only engage the platoon from the rear.

In Fig. 3, the transmission and reception of V2V and I2V (from Infrastructure) messages is schematically shown:

- Communication is performed from one truck to the next truck; bidirectional communication is established to control the platoon and the individual trucks in the platoon [12].
- The leading truck is in the lead of the platoon. Following trucks can provide 'constraints' to the platoon, e.g. as a result of their weight, their braking capacity, so that the leading truck can include that for the platooning cohesion and strategic decisions.
- A distinction is made between tactical platoon management messages (relevant for all platoon members and near-real time criticality) and platoon control messages (operational layer) that are required for controlling the individual trucks (acceleration-deceleration-cruising). The platoon control messages are of real-time nature, low-latency and mostly locally relevant: for the vehicle listening to the vehicle in front.
- A truck wanting to engage can transmit a Join Request to the Trailing truck; the Trailing truck will answer with a

Join Response with information on request acceptance and with connection and platoon configuration details.
- In ENSEMBLE also a security framework is used to set-up secure communications [13]. These mechanisms are out-of-scope for the current paper. These mechanisms are not relevant for our analyses and make the descriptions unnecessarily complex.
- Level A platooning allows for reception of information from the infrastructure, mostly strategic information such as advised following distance, traffic rules such as the speed limit and local traffic conditions. This information is intended to be received by each individual truck and by the platoon leader to take a decision.

## IV. SAFETY ASSESSMENT

A fundamental choice needs to be made for the safety assessment of platooning trucks, as there are two possible approaches:

### A. Assessment of the platoon as a system-of-vehicles

In the system-of-vehicles, the vehicles are cooperating following agreed rules on the use of V2X messages (the platooning protocol). This is the system boundary indicated by the orange bounding box in Fig. 3:
- Vehicles are able to receive messages from other vehicles in the platoon (V2V) and from the environment (I2V) and interpret these, like sensors that make an image of the environment as input to the decision and control of the vehicle.
- Transmission of messages is considered similar to the use of the actuators, such as steering, braking or gas pedal actuation. Input to the actuators is received from the decision and control algorithms. A transmitted V2V message is intended as input to the decision and control of one other vehicle in the platoon.
- The only information crossing the system boundary considers messages received from the infrastructure (this is one-way communication from the infrastructure to each truck) and the information exchange in case a new truck wants to participate in the platoon, and similar in case the last truck has decided to disengage from the platoon.

### B. Assessment of the response of the individual vehicle

In case individual vehicles are assessed, it is important to consider that the vehicles are part of a platoon, and that the vehicles respond also to input received by communication‡ (system boundary indicated by red bounding box in Fig. 3):
- Information is received from the vehicle in front and from the vehicle to the rear (V2V) for interpretation of decision and control software (very similar to a sensor).
- Moreover, a different type of information is received from the infrastructure, e.g. information on speed limit, desired time-headway, etc.

---

‡ The communication protocol described here results from the ENSEMBLE project; it describes in generic terms how trucks are agreed to communicate and interact. The platooning communication protocol is not standardized yet, but this work also contributes to ETSI TR 103 298: ITS; Platooning; Pre-standardization Study.

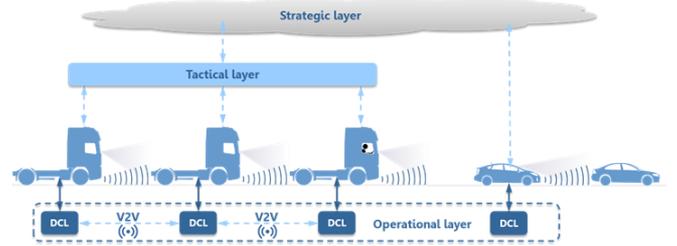

Fig. 4: Control layers used in ENSEMBLE truck platooning, where DCL represents the decision and control logic for an individual vehicle.

- Input from the decision and control software in the truck is used for the actuators and similarly to transmit operational messages. This is used by the vehicle to the rear for operational control (e.g. longitudinal vehicle following). This is typically part of the operational layer (see Fig. 4) and has a real-time nature. So this is V2V communication with low-latency and high update rates. The messages for platoon management are of a more tactical nature, so less time stringent, but of interest for all platoon members (upstream and downstream) to share status and coordinate platoon maneuvers.
- The information crossing the system boundary considers messages received from the infrastructure (one-way communication, tactical and/or strategic layer information), and the messages received from the preceding and following truck. Additionally the messages generated by the individual truck as a result of decision and control (similar to actuation) cross the system boundary. These outgoing messages only target the preceding truck or the following truck.

### C. Consider the role of the truck in the platoon

Although one might expect a result regarding the safety of a platoon as a whole when assessing the platooning use case, it is more meaningful to address the safety of an individual vehicle in the platoon. For Level A platooning, each individual vehicle remains responsible to behave safely, also as part of the platoon. Another fundamental argument results from the fact that we consider multi-brand platooning, which means that for each individual vehicle that has platooning functionality on board, the safety of this functionality needs to be assessed. Consequently, we propose to choose for assessing individual vehicles regarding its platooning functionality. The assessment needs to consider the different roles that a vehicle can have in a platoon (leader, follower, trailing), as its function changes with that role.

If we only would consider the platoon as a whole, then we would need to find a solution for the fact that platoons might have different compositions, with a different number of trucks, cooperation between different type of trucks and a different order of trucks in the platoon. In that case, safety assessment needs to be performed on all possible platoon configurations, which is simply not feasible.

## V. ADDING COMMUNICATION TO THE SCENARIO DESCRIPTION

The choice made in the previous section is relevant for the way scenario descriptions are impacted by adding communication. Now, from its definition, a scenario takes the perspective of the individual vehicle, which will not change. Actually, an additional information layer is needed for the scenario concept to deal with the information that is perceived (received) by the platooning vehicle. The assessment process moreover needs to consider whether the vehicle transmits appropriate signals (according to specification) based on provided input (also from V2V communication). Vehicle control and decision making by the individual truck depends on the V2V and I2V inputs that the truck receives. Moreover, often also the content of the transmitted V2V messages of the individual truck, depends on the information previously received by the individual truck.

In Fig. 4, the three possible roles for a platooning truck are indicated. A fourth role is a platoon candidate, that is a truck that wants to engage into the platoon (from the rear in case of Level A platooning). It appears that for platooning it is important to indicate the position of the truck in the platoon. When the specific functionality of a vehicle is coupled to a scenario, this is called a use case. So in the description of the use case "Platooning", it needs to be indicated what the role of the truck in the platoon is. Based on the role, a selection of applicable scenario categories can be made. In the description of the scenario, then also the communication layer is considered. The communication input signals are part of the scenario, and scenario evolution, the transmitted V2V messages of the truck – as part of the total response of the truck, is part of the performance assessment. The information flow to and from individual trucks in the platoon depending on their role has been schematically shown in Fig. 5 and Fig. 6.

### A. Example scenario

We use the example of a 3-vehicle platoon consisting of a lead truck, a following truck and a trailing truck, running into the tail of a traffic jam, to explain how the existing scenario information needs to be extended with communication information (in *italics*). We start the scenario description providing the initial state from which the scenario commences:

- The platoon is driving in the right lane of a 3-lane highway with a speed of 80 km/h. In front of the leading truck is a passenger car (at 80 km/h), at a given time-headway.
- The road is not equipped with display panels at gantries over the road that warn for a traffic jam, indicating an appropriate speed limit and using flashing amber lights next to the display panels for additional warning.
- The scenario commences as the passenger car preceding the leading truck starts to brake with 3 m/s2 in anticipation of the upcoming traffic jam.

Subsequently, we will provide the scenario descriptions for each of the trucks, considering the role of the truck in the platoon. The input-output schemes for the different roles in the platoon are provided in Fig. 7 and Fig. 8.

### B. Leading truck

The Platoon Leader is driven by a human driver. The driver is supported by an Advanced Driver Assistance System that uses a sensor system to get a view on the traffic in front and to the side of the truck, so the Platoon Leader is in ACC mode. Sensor inputs are used to build and continuously update a world model that the Automated Driving System (with platooning functionality) of the leading truck uses to support the driver, to adapt the speed, acceleration and steering angle and to communicate its intentions to the following trucks. In its decisions, the leading truck also makes use of information it receives from the following trucks and from I2V. For the current example, we assume no additional information from I2V to be present.

*1) Intention of the leading truck*
- Safely decrease speed to a level that is appropriate for the traffic jam with a deceleration that does not exceed that of the least performing[§] truck in the platoon. For this it is assumed that the driver of the Platoon Leader is aware of the traffic jam by visual observation.
- *Communicate this intention and the according braking deceleration and speed or gap (THW) adjustment to the following trucks with the shortest possible delay.*
- Continuously monitor the deceleration profile of the preceding passenger car, the THW and the TTC (time-to-collision).
- *Communicate the reason for speed or gap adjustments downstream into the platoon as a result of the current scenario (approaching the tail of a traffic jam).*

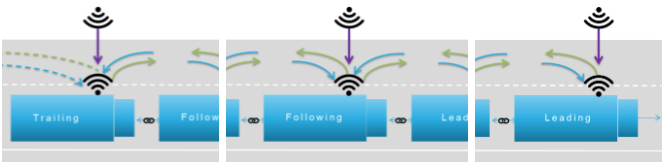

Fig. 5: System boundaries for the Trailing truck, the Following truck and the Leading truck.

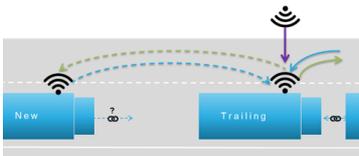

Fig. 6: Role of a "new" candidate truck asking permission to engage in the platoon.

---

[§] Least performing in the sense of deceleration profile which is the result of the braking capability in combination with the payload of each truck.

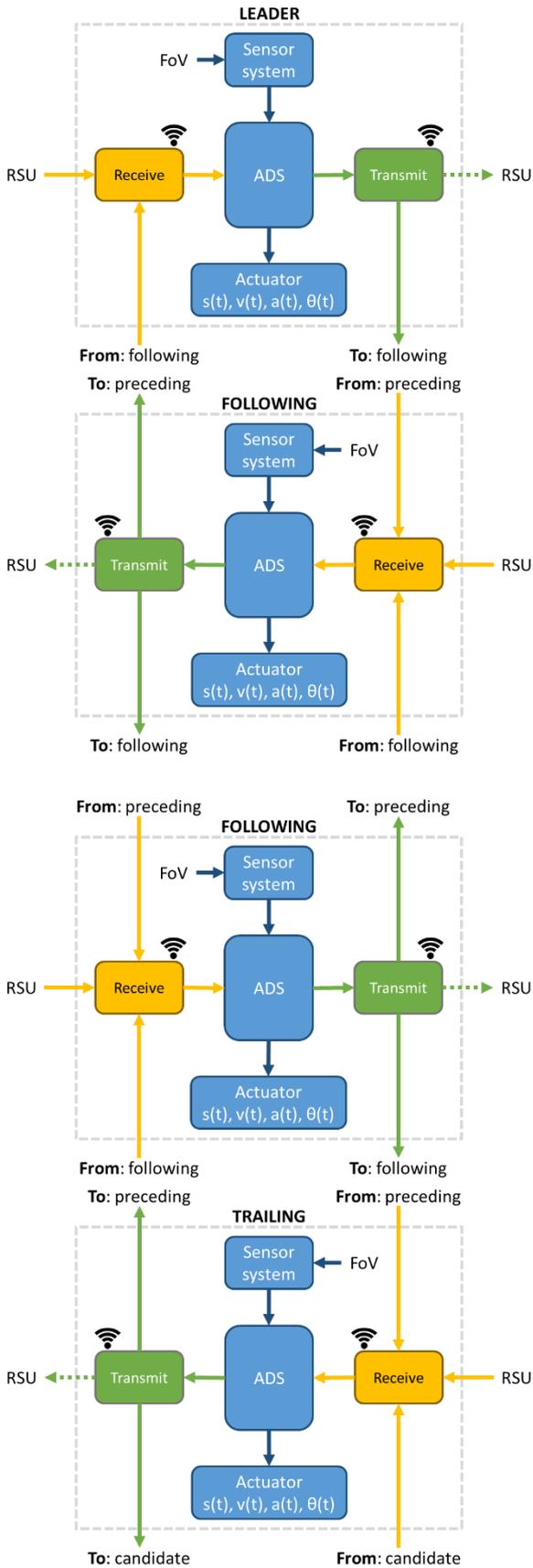

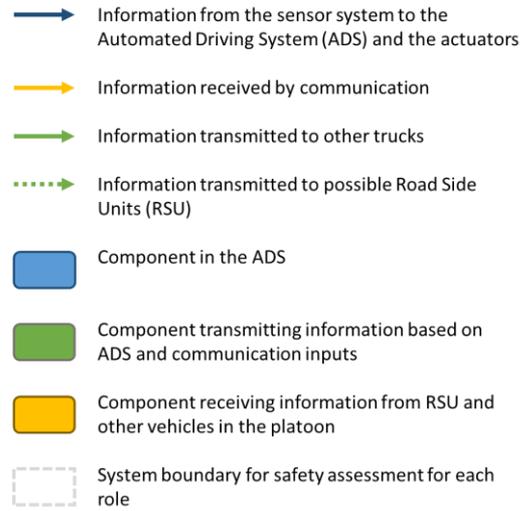

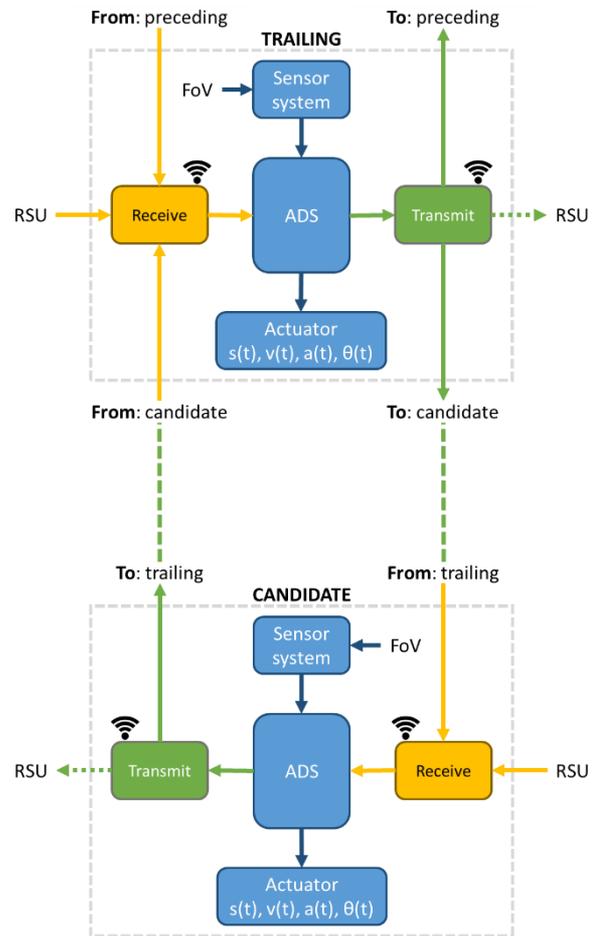

Fig. 7: Input-output scheme for the different roles in a platoon: for leading, following and trailing truck.

Fig. 8: Input-output scheme for new trucks with the intention to engage into the platoon.

*2) Driver supporting sensor system of the leading truck*
- The leading truck notices the preceding vehicle braking at a distance of 300 m to the tail of the traffic jam.
- The traffic jam covers all three lanes, the last vehicles in the queue use flashing blinkers to indicate a strong reduction in speed. The actual traffic jam consists of slow driving and often standing still vehicles.

*3) Communication regarding the leading truck*
- The leading truck is aware (due to earlier communication during platoon engagement) of the platoon state regarding the number of trucks, the THW between the trucks and the max. braking capability of the least performing truck in the platoon.
- There is no I2V information, so no additional information is received from the traffic management center (such as warning for a traffic jam – minutes before actually running into it, or speed advice.)

### C. Following truck

The first following truck is platooning at 1.5 s THW to the leading truck. The trailing truck is platooning at 1.5 s THW from the first following truck. The platooning functionality uses longitudinal control only, to keep the THW to the preceding truck constant; a human driver controls the steering wheel to keep the following truck into its lane. The sensor system monitors the preceding truck, and also checks for unexpected maneuvers of other traffic participants, e.g. to anticipate on a vehicle cutting through the platoon in between the leading truck and the following truck – which would change the scenario for the following truck(s).

*1) Intention of the following truck*
- Keep the THW with respect to the leading truck.
- Receive and follow the deceleration and speed profile that the leading truck has communicated down the platoon.
- Continuously monitor the THW and the TTC (time-to-collision) with respect to the leading truck and adapt the deceleration where necessary.
- *Communicate ego intentions and related platoon cohesion information to the leading truck and the trailing truck with the shortest possible delay.*
- *Receive platoon cohesion information of the trailing truck to inform the leading truck.*

*2) Sensor system of the following truck*
- The following truck monitors the position and speed of the leading truck.
- In addition, the sensor system monitors the road users that possibly interfere with the platoon, e.g. by cutting through the platoon in between the leading and the following truck.

*3) Communication regarding the following truck*
- The following truck receives information from the leading truck on the requested deceleration and speed profile or gap adaptations for the platoon.
- It also receives platoon cohesion information from the trailing truck e.g. desired maximum platoon speed, maximum acceleration request.
- There is no I2V considered, so no additional information is received from the traffic management center.

### D. Trailing truck

The trailing truck is platooning at 1.5 s THW from the first following truck. The platooning functionality uses longitudinal control only, to keep the THW to the preceding truck constant; a human driver controls the steering wheel to keep the following truck into its lane. The sensor system monitors the preceding truck, and also checks for unexpected maneuvers of other traffic participants, e.g. to anticipate on a vehicle cutting through the platoon in between the following truck and the trailing truck – which would change the scenario for the trailing truck.

*1) Intention of the trailing truck*
- Keep the THW with respect to the truck in front.
- Receive and follow the deceleration and speed profile that the leading truck has communicated down the platoon.
- Continuously monitor the THW and the TTC (time-to-collision) with respect to the following truck and adapt the deceleration where necessary.
- Communicate this intention and related platoon cohesion information to the following truck.

*2) Sensor system of the trailing truck*
- The trailing truck monitors the position and speed of the following truck.
- In addition, the sensor system monitors the road users that possibly interfere with the platoon, e.g. by cutting through the platoon in between the following and the trailing truck.

*3) Communication regarding the trailing truck*
- The trailing truck receives information from the preceding truck on the requested deceleration profile.
- It also receives information from a Platoon Candidate, a new truck that expresses its intention to join the platoon, incl. its max. braking capability.
- A decision on whether or not to allow a Platoon Candidate to engage is transmitted to the Platoon Candidate via a Join Response message.
- If a Platoon Candidate is allowed to Join, additional information is shared about platoon configuration: Platoon ID, communication configuration, and number of trucks.
- When the engage has finished successfully, related platoon configurations are updated accordingly (e.g. platoon roles, position in the platoon, number of vehicles in platoon). This information is aggregated and forwarded from rear to front.
- There is no I2V, so no additional information is received from the traffic management center.

*E. Platoon candidate*

In case a Platoon Candidate intends join the platoon, it transmits a Join Request to the trailing truck of the platoon. This request is handled locally by the trailing truck as it has all the information available to decide whether or not to allow the new truck to engage from the rear based on a set of requirements, such as:
- The maximum number of trucks in the platoon should not have been reached.
- The new truck should be able to communicate according to the same protocol as the trucks in the platoon.
- The maximum deceleration capability of the new truck matches that of the already participating trucks in the platoon sufficiently.
- The situation allows for the engaging maneuvers to be conducted safely without adding hazard to the platoon or other road users.

## VI. DISCUSSION

The example shows how a scenario translates into input to the cooperative automated driving system (CADS), by the sensor system, by V2X communication, and by the driver (for the leading truck, as well as by the safety driver in the other trucks in the platoon). The decision and control logic of the CADS on-board each vehicle in the platoon interprets the incoming signals and continuously provides a response. This response results in control of the actuators, the transmission of information across the platoon, and information on an HMI to the driver. A scenario for a given truck in the platoon is described by indicating the intention of the truck (based on its role), the (dynamic) behaviour of the road users in the direct environment of the truck (which obviously includes the behaviour of the other truck(s) in the platoon whenever relevant) as monitored by the sensor system, the received information through V2V and I2V communication. For the safety assessment of a truck in a platoon, this would lead to at least the following tests:

1. ***Sensor perception tests***: how well do the sensors in the sensor set perceive the maneuvers of other road users and objects in the infrastructure (lines, lane markings, traffic signs, etc.), and how accurate is the resulting world model for different lighting and weather conditions?

2. ***Communication (V2X) tests***: what is the quality of the communication, is the channel uncongested, are V2V messages delayed, or lost, from preceding and following vehicles? How does this quality depend on the scenario, e.g. by influences from its environments: infrastructure elements such as tunnels, bridges, or gantries, or from weather conditions?

3. ***Open loop tests***: what is the output of the CADS in response to the various inputs? In open loop tests, the set points for control of the actuators are determined and recorded, but not used for actual actuation. Moreover, the messages for broadcasting information up and down the platoon are formulated and recorded for an offline consistency check.

4. ***Closed-loop tests***: these are the tests in which the set points determined by the CADS are used to control the vehicle. This includes the transmission of information, and using the received information from V2V communication in the determination of set points and to update outgoing V2V messages.

A set of open-loop and closed-loop tests is required for each role that a truck may play in the platoon. Such tests should also evaluate the role-awareness of each individual truck in the platoon.

## VII. CONCLUSION

For the safety assessment of trucks with V2V platooning functionality, a methodology has been proposed that considers the role of the individual truck in the platoon. In this way, each individual truck is tested for its role, given a set of test cases based on scenarios that not only describe the intent of the vehicle-under-test, the static & dynamic environment and the environmental conditions, but also the messages that are communicated between the vehicles (V2V communication) and between infrastructure and the vehicle-under-test (V2I). Since the communication layer is added to the scenario descriptions, the scenarios have become dependent on the chosen communication protocol for platooning. In the paper the protocol according to the ENSEMBLE project is followed.

As the next step, the set of relevant test cases needs to be collected, e.g. using a StreetWise approach [14]. For each of the relevant highway scenarios the communication layer needs to be added referring to the role of the truck-under-assessment and according to the selected communication protocol.

This approach enables the safety assessment of platoons by assessing the individual trucks. Close-loop tests can be performed either in a virtual model environment, on a test track, or in a X-in-the-loop environment, and the open-loop tests might even be performed on the public road.